\documentclass[preprint,journal]{vgtc}       % preprint (journal style)

\ifpdf%                                % if we use pdflatex
  \pdfoutput=1\relax                   % create PDFs from pdfLaTeX
  \usepackage{graphicx}                % allow us to embed graphics files
  \DeclareGraphicsExtensions{.pdf,.png,.jpg,.jpeg} % for pdflatex we expect .pdf, .png, or .jpg files
\else%                                 % else we use pure latex
  \usepackage{graphicx}                % allow us to embed graphics files
  \DeclareGraphicsExtensions{.eps}     % for pure latex we expect eps files
\fi%

\graphicspath{{figures/}{pictures/}{images/}{./}} % where to search for the images

\usepackage{microtype}                 % use micro-typography (slightly more compact, better to read)
\PassOptionsToPackage{warn}{textcomp}  % to address font issues with \textrightarrow
\usepackage{textcomp}                  % use better special symbols
\usepackage{mathptmx}                  % use matching math font
\usepackage{times}                     % we use Times as the main font
\usepackage{cite}                      % needed to automatically sort the references
\usepackage{tabu}                      % only used for the table example
\usepackage{booktabs}                  % only used for the table example
\usepackage{enumitem}
\usepackage{gensymb}

\usepackage{color,soul}
\onlineid{1076}

\vgtccategory{Research}
\vgtcpapertype{please specify}

\title{Sea of Genes:\\ A Reflection on Visualising Metagenomic Data for Museums}

\author{Keshav Dasu, \textit{Student Member, IEEE}, Kwan-Liu Ma, \textit{Fellow, IEEE}, Joyce Ma, and Jennifer Frazier}
\authorfooter{
\item
 Keshav Dasu is with UC Davis. E-mail: kdasu@ucdavis.edu.
\item
 Kwan-Liu Ma is with UC Davis. E-mail: ma@ucdavis.edu.
\item
 Joyce Ma is with The Exploratorium. E-mail:  jma@exploratorium.edu.
\item
 Jennifer Frazier is with The Exploratorium. E-mail: jfrazier@exploratorium.edu.
}

\shortauthortitle{Biv \MakeLowercase{\textit{et al.}}: Global Illumination for Fun and Profit}
\abstract{We examine the process of designing an exhibit to communicate scientific findings from a complex dataset and unfamiliar domain to the public in a science museum. 
Our exhibit sought to communicate new lessons based on scientific findings from the domain of metagenomics. 
This multi-user exhibit had three goals: (1) to inform the public about microbial communities and their daily cycles; (2) to link microbes' activity to the concept of gene expression; (3) and to highlight scientists' use of gene expression data to understand the role of microbes.
To address these three goals, we derived visualization designs with three corresponding stories, each corresponding to a goal. 
We present  three successive rounds of design and evaluation of our attempts to convey these goals. 
We could successfully present one story but had limited success with our second and third goals. 
This work presents a detailed account of an attempt to explain tightly coupled relationships through storytelling and animation in a multi-user, informal learning environment to a public with varying prior knowledge on the domain and
identify lessons
for future design.

} % No line break to make sure there is no space after the em-dash after "Abstract".


\CCScatlist{ % not used in journal version
 \CCScat{K.6.1}{Management of Computing and Information Systems}%
{Project and People Management}{Life Cycle};
 \CCScat{K.7.m}{The Computing Profession}{Miscellaneous}{Ethics}
}

\vgtcinsertpkg

\firstsection{Introduction}

\begin{document}

\maketitle
% \IEEEraisesectionheading{\section{Introduction}}
Visualizations are increasingly central to the practice of science.
They are used across a range of scientific disciplines to analyze phenomena, such as changes in microbiomes and shifts in climate.
% As more visualizations are created by scientists, they have been increasingly present in science museums. 
% In the last decade, there have been several significant efforts to bring scientific and information visualizations to science museums.
% for innovating the means of design leading to successful delivery.
% Visualizations provide museums with the means to give the public an understanding of scientific insights.
There have been several large-scale efforts to develop scientific and information visualizations for the public: the National Oceanic and Atmospheric Administration's (NOAA's) Science on a Sphere presents earth systems datasets such as tsunamis, climate models, and sea surface temperature on a large spherical display for aquariums and museums~\cite{SOS}; DeepTree~\cite{DeepTree} visualizes evolutionary data for exploration on a tabletop interface in natural history museums; MacroScope~\cite{smit2018macroscope} ports a range of visualizations into a large interactive display for a wide range of academic and museum settings; and Living Liquid~\cite{JMa2012,hsueh2016fostering} created interactive visualizations for a hands-on museum environment. 
Each of these projects, as well as many others~\cite{tGeller2006,schmidt2007,hinrichs2011}, have contributed to our understanding of the opportunities and limitations of visualizations in museum settings. 
% All of these projects, however, visualized concepts, such as currents, weather, evolutionary trees, or migration paths that are somewhat familiar to the public.
However, these projects visualized concepts such as currents, weather, evolutionary trees, and migration paths that the public has familiarity with.

\begin{figure}[h]
	\includegraphics[width=\columnwidth]{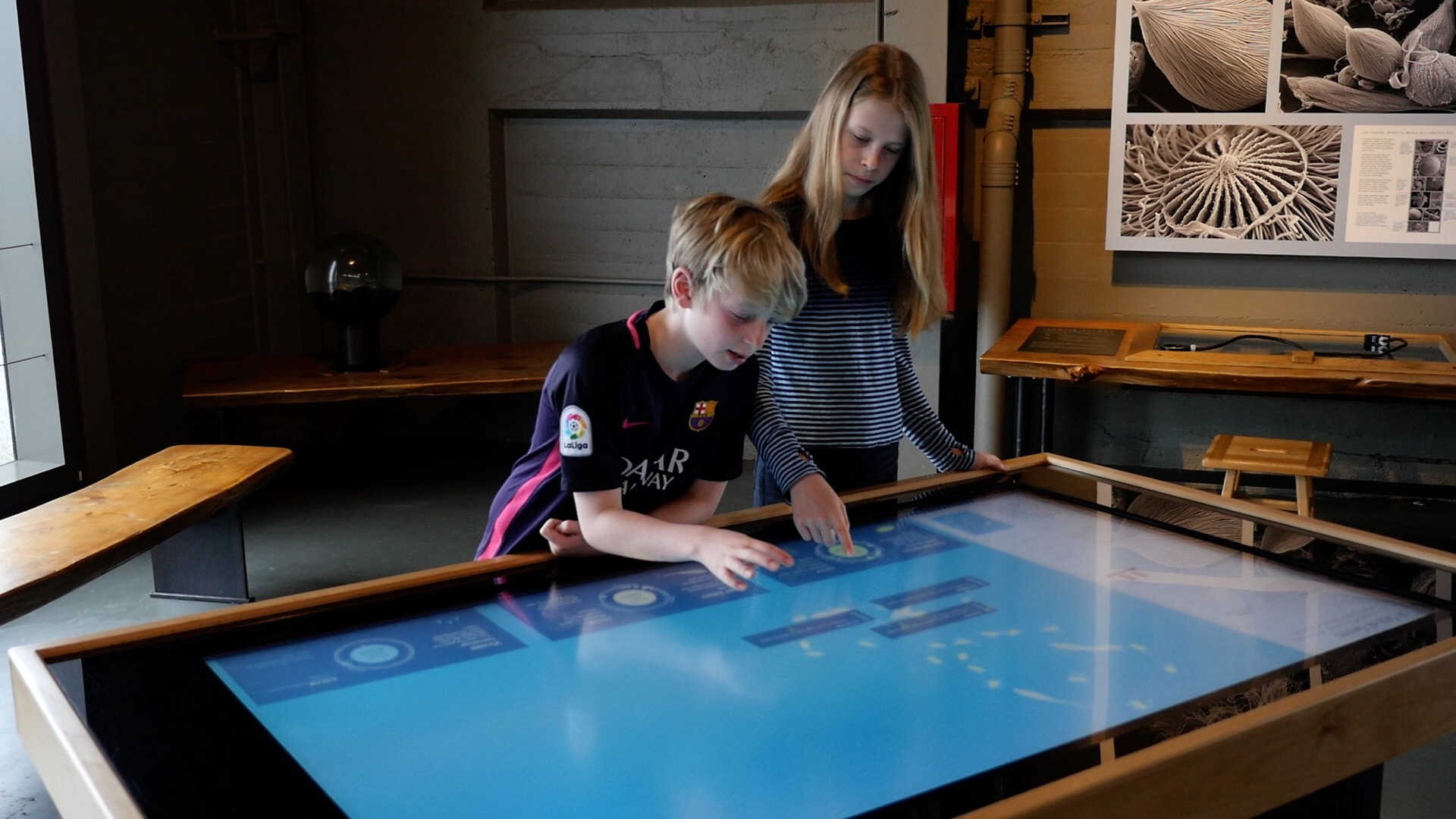}
  	\caption{Museum visitors using the Sea of Genes exhibit in the Life Sciences Gallery of the Exploratorium in San Francisco. }
  	\vspace{-3mm}
	\label{fig:visitors}
\end{figure}

This paper examines the challenges of creating a museum exhibit from a complex dataset from an emerging and unfamiliar field: metagenomics.
Metagenomics, the characterization of all the genetic data in a sample, is revolutionizing our understanding of microbes. 
% The data collected contains information such as the gene expressed and the peak time of expression for these microbes.
Researchers use these data to determine what species are present, what functions they perform, how these functions change over time, and infer how microbes interact~\cite{national2007new}. 
% tringe2005comparative
Metagenomics is one of the primary ways researchers study microbes.
Microbes play a central role in almost all aspects of life on earth~\cite{national2007new}.
Ocean microbes use energy from the sun to produce half the oxygen we breathe and drive our climate; soil microbes impact the food we eat; and scientists are beginning to understand the complex and critical roles billions of microbes living in our bodies have on our health~\cite{national2007new}.
Despite the significance of metagenomics to scientific research, few efforts have introduced these data to the public through interactive visualization.
Instead, exhibits~\cite{SecretWorld,InvisibleYou,Zoo} rely on electron micrographs and graphics of microbes.
% , and artist pieces.

Even though the visualization field has explored narrative elements as a strategy for engaging users with complex data~\cite{segel2010narrative,Chan2016,Bryan2017}, there has been limited work on visualization exhibits that present complex content from an unfamiliar domain in a museum setting.
Unfamiliar domain, in this context, refers to the targeted audience having little prior knowledge of the domain.
We examine the application of narrative visualization strategies and animation effects to reduce complexity and create familiarity when presenting scientific findings through a museum exhibit.
The exhibit is called \textit{Sea of Genes}.
% We explore how the effectiveness of animation when the instructional domain of the content to be learned lacks familiarity. 
First, we provide a detailed documentation of the design process for developing such an exhibit and the challenges we faced compared to our prior exhibit design experiences.
We then address limitations and constraints of designing a visualization exhibit in an informal learning environment and point out directions for research in this space.
% In particular, our study examines the combination of animation and narrative strategies and how they impacted visitors’ interpretations and understandings of metagenomic data.
This work presents a detailed account of an attempt to explain tightly-coupled relationships through storytelling and animation in a multi-user, informal learning environment to the public with varying prior knowledge on the domain.
We discuss takeaways and provide guidance for studying science museum exhibits, which we believe is especially valuable to both the field of metagenomics and other scientific domains.
% and contribute to lessons learned in communicating the research of

\section{Related Work}
%\section{Animation}

%This section focuses on research on the use of animation for teaching unfamiliar concepts and effective design principles.
%illustrate what we from animation we are evaluating in this unfamiliar context

Extensive work has been done on the application of narrative devices and visualization of complex data.
There has also been research on the use of animation for teaching unfamiliar concepts.
With \textit{Sea of Genes} we paired principles from both educational animation and narrative visualization to develop an exhibit that could successfully present content from both a complex dataset in an unfamiliar domain.

\subsection{Value of Narratives}
Kosara and Mackinlay~\cite{kosara2013storytelling} define a narrative as ``a clearly defined path that is composed of a sequence of ordered steps, containing either text, images, visualizations, video, or any combination of the latter''.
Each step can be thought of as a story that relates the `who, where, when and how' of an event that occurred.
The ordering of these steps creates the narrative.
How the order is set impacts its reception.
% How the order is set impacts how it is received.
Storytelling is the activity of sharing a narrative.
Theorists posit narratives are fundamental to human sensemaking~\cite{bruner2009actual} and intelligence~\cite{schank1995tell}.
Research~\cite{dudukovic2004telling} has found stories help us connect and remember facts. 
% Written narratives are easier to understand and recall than expository text \cite{graesser2002does}. 
The roles and forms narratives may serve and can take in visualization continue to be active areas of research. 
For example, Gershon and Page~\cite{gershon2001storytelling} discuss the value of storytelling when developing applications. 
% These environments process streams of information and data sources possibly in real time. 
Segel and Heer~\cite{segel2010narrative} provide a comprehensive review of narrative visualization as used by online journalists.
These works motivate the value of using narrative to communicate concepts to the lay audience.

%
% Storytelling and Narrative in Vis
%
\subsection{Narrative Frameworks}
Many visualization researchers have analyzed and designed a number of methodologies and frameworks to apply narrative techniques~\cite{segel2010narrative,kosara2013storytelling,gershon2001storytelling,hullman2011visualization,ma2011scientific}. 
Narratives convey a message, enhance comprehension, make transparent causality, increase engagement, and summarize and simplify a complex message. 
The narrative visualization literature describes a number of approaches on how to embed a narrative.
% These different frameworks seem to be driven to an extent by the motivation for storytelling. 
Segel and Heer’s case study of Gapminder Human Development Trends~\cite{segel2010narrative} explores how narratives enable complex information to be comprehended quickly by the user.
% This is achieved through the use of graphical elements such as animations and the use of color schemes. 
Hullman and Diakopoulos~\cite{hullman2011visualization} discuss a set of omission techniques used for simplifying complex data. 
Lee et al.~\cite{lee2015more} describe a process of working with data analysts to extract only what is relevant to the story as well as how setting and device choice influence the presentation of a story.
% All three studies present different approaches for presenting complex data.
These works provided insights for~\textit{Sea of Genes} on how to use narrative to communicate lessons from the complex metagenomic dataset.
We expand on the current literature with how the museum setting, especially a highly interactive one, affects the presentation of a narrative.
% With Sea of Genes the content is both complex and unfamiliar so we applied each of these approaches in an effort to communicate our narrative.

% Since narratives make sense of facts, it follows that they can also influence how data is perceived. 
% Hullman and Diakopoulos \cite{hullman2011visualization} discuss how omissions can also deliberately lead to bias. 
% Similarly, biases can also result from over-emphasis. 
% Based on concepts from critical theory, semiotics, journalism, and political theory, they present a comprehensive list of techniques for shaping the visual narrative to convey a targeted message.
Gershon and Page \cite{gershon2001storytelling} contend conveying a story in general is more effective when images are combined with text or data. 
While images in themselves convey a lot of information, we need data and text to reduce the ambiguity in the message. 
They also identify continuity as an important element.
Continuity may result from a causal flow that also enables retention and recall. 
If a user perceives continuity, this may also imply they have understood the causality.
% Continuity in this context is linked to sequence wi
Our design must focus on preserving the continuity between each step in our narrative, since the sequence in which they are presented matters for our stories.
% Therefore we explored methods to aid users in perceiving continuity.

Segel and Heer \cite{segel2010narrative} provide an overview of how visual elements have been employed in traditional media such as comics, books, and films to tell stories. 
Their focus is on the role of graphical elements and interactivity in maintaining continuity in the flow of the narrative. 
They identify author-driven and reader-driven as two polar extremes of visualization.
We contextualize these two terms for the museum setting as designer-driven and visitor-driven.
In a designer-driven approach, the story is linear and the visitor has no control of the narrative.
It presents to the visitor a fixed sequence of events with which they can interact.
% That is, the visitor is presented a fixed sequence of events with which they can interact and view. 
In a ``pure'' visitor-driven approach, there is no predefined narrative.
Instead, there is no fixed sequence of events, and the visitor would select and order events to create a narrative.
% They are concerned with approaches for balancing these two extreme approaches.
% Insert citation about the work that is concerned with telling this approach
% They identify three techniques for blending these two approaches: martini glass, interactive slide show, and drill down story. 
By blending these two extremes in our own design, we seek to retain continuity and provide visitors freedom to explore.

% According to Segel and Heer and Hullman and Diakopoulos, continuity, which is viewed as being central to story telling~\cite{segel2010narrative,hullman2011visualization}, is needed to communicate a message. 
% While interactivity is needed for exploration and learning.
Narratives have been widely applied in history and art museums to help visitors make personal connections to an object or a collection~\cite{Bedford,Rounds2002}. 
% To a lesser extent, narrative has been incorporated into natural history and science museum exhibits and exhibitions with mixed results.  
A study~\cite{FindingSig} on the roles narratives play in interactive science exhibits found enhancing exhibits with personal stories improved the exhibit experience for visitors and helped create personal connections to the content.
However, adding stories seemed to reduce the visitors' physical interactions and explorations with the exhibit. 
Similarly, a study~\cite{ma2013engaging} examining the use of narrative introduction to describe the dataset visualized in an exhibit, found it did not improve data exploration. 
% When and how to incorporate narrative into science exhibits remain unanswered design questions in the museum field.
% The effectiveness and applicability of narrative for communicating content that is both complex and unfamiliar, when applied to an interactive visualization exhibit, requires further study.
Further study of narrative applied to interactive visualization is needed, examining its applicability and effectiveness in communicating complex and unfamiliar content.
We identified a set of related stories, which we could present in layers.
\textit{Sea of Genes} provided us with an opportunity to assess the effectiveness of applying the aforementioned narrative strategies to breakdown a complex metagenomic dataset to then present a layered story to the public.
% The previous work suggests that by using these techniques we can explain the complex metagenomic content to novices.

\subsection{Animation for Learners}
Research on how animation affects learning has gone through two eras of consideration.
In the first era (1990's), researchers studied the impact animation has on learning by evaluating it next to static graphics \cite{tversky2002animation, hegarty2003roles,moreno2007interactive}.
These studies report inconsistent or inconclusive findings on the effects of animation on learning.
In particular, although Schnotz and Grzondziel~\cite{schnotz1999individual} found animation performed better, it had an interactive component~\cite{ferguson1995learning} confounding the results.
Tversky and Morrison~\cite{tversky2002animation} were highly skeptical animations could be effective for conveying complex systems.
They suggest two principles to note as conditions for an animation to be effective: Apprehension and Congruence. 
The \textit{Apprehension} principle states ``the structure and content of the external representation should be readily and accurately perceived.''
A drawback of animation is the perceptual and cognitive limitation of processing a changing visualization, e.g. complex interactions may occur too quickly to be understood.
The \textit{Congruence} principle states ``the structure and content of the external representation should correspond to the desired structure and content of the internal representation.''
In principle, animation should be effective for expressing changes.
Most animations violate these principles. 
% Some have too many moving parts making it difficult for a viewer to know what to observe.
People conceive a dynamic process as a sequence of steps, thus violating the Congruence principle.
In order for an animation to be effective Tversky~\cite{tversky2002animation} believes animations must explain rather than simply show.

Rather than comparing animations' effectiveness to static graphics, recent studies focus on understanding the cognitive processes involved in processing dynamic visualizations and identifying the steps leading to comprehension \cite{morrison2000animation, ainsworth2008animations}.
Berney and Betrancourt~\cite{berney2016does} conduct a meta-analysis on animation for learning and section the factors into three main groups: (a) specific to the learners, (b) specific to the instructional material, and (c) specific to the learning context.
% One constraint of Sea of Genes design is that the domain is unfamiliar to the visitors. 
% This stems from visitor's prior knowledge of this domain varying widely from most having limited to no knowledge. 
Studies~\cite{chanlin1998animation,kalyuga2008relative} that address group (a), to which museum visitors belong, found varying prior knowledge requires varying presentation forms to achieve a learning task.
Therefore, with \textit{Sea of Genes} we need to consider other ways to enforce the animation, which we detail in Section 5.
% Since our visitors' prior knowledge of the domain vary widely, due to the domain being unfamiliar, we needed to consider other ways to reinforce the animation.
Chanlin \cite{chanlin1998animation} found animation enhanced both novice and experienced learners' learning.
Specifically, for novices it helped facilitate learning of descriptive facts.
Berney and Betrancourt form a hypothesis based on Ploetzner and Lowe's~\cite{ploetzner2012systematic} work that ``well-designed'' expository animations contain all the elements needed to draw learners' attention to the right place at the right time.
That is, our animation should facilitate directing visitors attention to each point when necessary.

\section{Sea of Genes }
\begin{figure*}
	\includegraphics[width=0.95\textwidth]{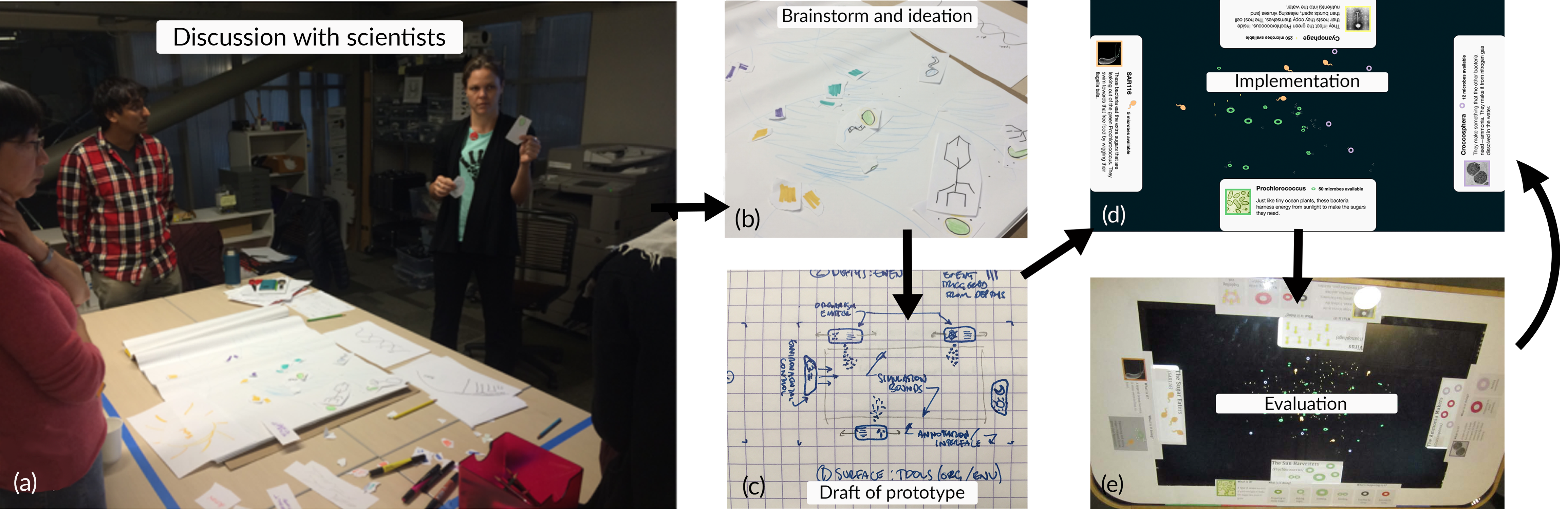}
	\centering
  	\caption{Design process (a) Photo capturing a discussion from the initial brainstorm session with a scientist from C-MORE. (b) Sketches of microbe behavior inferred from genomic data at the brainstorm. (c) At the end of the brainstorm we adpoted a sketch of Prototype 1 produced by Stamen Design. (d) Prototype 1 still. (e) Prototype 1 on the floor of the Exploratorium during evaluation. 
%   	\vspace*{-0.12in}
  	\vspace{-3mm}
  	}
	\label{fig:brainstorm}
% 	\vspace{-5mm}
\end{figure*}

\textit{Sea of Genes} is a multi-user interactive visualization exhibit at the Exploratorium, a science museum.
The exhibit is within the Living Systems Gallery of the museum, and uses a combination of animation techniques and narrative elements to communicate three key stories found in a metagenomic dataset of marine microbes:
\begin{itemize}
\setlength{\topsep}{-0.03in}
\setlength{\itemsep}{-0.03in}
    \item Microbial interactions occur in a predictable daily rhythm.
    \item Genes turn on and off according to a daily rhythm.
    \item Scientists collect data about the genes of microbes.
\end{itemize}
This exhibit was created through an interdisciplinary collaboration that brought together expertise in academic and commercial visualization practices, scientific research, and museum design and evaluation.
The Exploratorium led the collaboration and provided a curator, project manager, writer, graphic design, exhibit designer, learning researcher, and evaluators.
A visualization group provided a graduate student and professor to provide expertise in visualization and HCI research to assist the exhibit design.
The University of Hawaii at Manoa provided a lead data scientist and marine microbiologist who provided datasets and content expertise.  
Stamen Design provided a digital graphic designer and visualization designer to provide expertise in public-facing commercial visualization design and public installation.  
% This collaboration not only served to create the exhibit, but was also a vehicle for professional development of the team members in educational visualization.
% The exhibit was created through a collaboration between a marine microbiology research group, a visualization research group, the Exploratorium, and Stamen Design.
% The marine microbiology group provided data and content expertise; the university visualization group brought expertise in HCI and visualization research; the Exploratorium brought expertise in exhibit design and evaluation; Stamen Design brought expertise in graphic design, techincal skill, and visualization insight from industry.
% Through this collaboration, we designed visualizations using a combination of animation techniques and narrative elements to communicate three key stories found in a metagenomic dataset of marine microbes:

\subsection{The Dataset}
The data used for \textit{Sea of Genes} were collected and analyzed by oceanographers affiliated with the Center for Microbial Oceanography: Research and Education (C-MORE) at the University of Hawai’i at Manoa and the Monterey Bay Aquarium Research Institute (MBARI).
A full description of the data collection and analysis methods were published in a series of articles~\cite{ottesen2014multispecies, aylward2015microbial, aylward2017diel} during 2014--2017. 

The 2014--2015 samples were collected using an Environmental Sampling Processor (ESP)~\cite{ottesen2014multispecies, aylward2015microbial}, a free-drifting sampling device that collects environmental and genomic data at specified times in the ocean, in this case, every 4 hours for 3 days. 
The 2017 samples were collected every 4 hours for 4 days using Niskin bottles~\cite{aylward2017diel} deployed from a research vessel. 
Planktonic microbial assemblages were collected by passing seawater through a 0.22 $\mu$m pore-sized filter, preserved in RNA later, and stored at -80\degree C within 24 hours of retrieval from the instrument.
RNA was extracted, cDNA was generated, and Illumina sequencing~\cite{aylward2017diel} was performed. 
Metatranscriptome reads were mapped to ortholog clusters of proteins constructed from the phylogenetic groups of interest. 
Function was assigned by KEGG Orthology annotation. 
Read count tables were normalized to total read count, with threshold set to achieve R2 value $>$ 0.8 using the R packages igraph and WGCNA~\cite{langfelder2008wgcna}. 
These count tables contained information about daily patterns in microbes such as: time of collection, taxonomic assignment, gene function and expression levels, and the peak time of expression. 
From this, C-MORE and MBARI scientists were able to infer which microbes were present, what functions they performed, and when those functions occurred over the course of a 24 hour period. 
They provided us access to these data sets and assisted in interpretation, (Figure~\ref{fig:brainstorm}a).

% The data used for Sea of Genes was collected and analyzed by ocean researchers at the Simons Collaboration on Ocean Processes and Ecology (SCOPE) at the University of Hawaii.  
% They were collected using an Environmental Sampling Processor (ESP)~\cite{aylward2015microbial}, a glider that collects environmental and genomic data at specified depths in the ocean.
% Planktonic microbial assemblages were collected in the 0.225 microgram size range at 4-hour intervals and preserved aboard the ESP. 
% Filter samples collected by the ESPs were stored at \ang{-80}C within 24 hours of their retrieval. 
% RNA was extracted, cDNA was generated, and Illumina sequencing was performed. 
% Metatranscriptome reads were mapped to ortholog clusters of proteins constructed from the phylogenetic groups of interest, and a weighted transcriptomic network approach using the R packages WGCNA~\cite{WGCNA} and igraph were used to analyze the resulting count tables. 
% The resulting data contain information about microbes such as: the depth of collection, gene expressed, and the peak time of expression.
% From this, SCOPE scientists infer which microbes are present and what functions they are performing.
% They provided access to parts of the dataset and assisted in interpretation, Figure~\ref{fig:brainstorm}a. 
% These data were the basis for the 2014 study published from University of Hawaii in the journal Science~\cite{ottesen2014multispecies}. 
% The scientists collected samples by using an Environmental Sample Processor (ESP) suspended at a specified depth. 

\subsection{Hardware and Implementation}
% \KLMC{***I think more should be said about selection of the table and the implementation.}\\
For \textit{Sea of Genes} we decided to use an interactive touch table, which has been shown to encourage collaboration and attract attention~\cite{hornecker2008don,rogerCollab2004,potvin2012comparing}. 
% jacucci2010worlds
A feature of the museum context is the ability to support social experiences~\cite{falk2000learning}.
Hinrichs et al.'s \cite{hinrichs2008emdialog} findings suggest using a large interactive display gives the visualization a presence within an exhibition.
These displays allow people to enjoy and participate from a distance and decide whether to engage further.
In our previous project we used the Multitaction object-tracking table \cite{JMa2012}, which attracted and engaged visitors with the visualization.
We took advantage of the social context of the museum by using a larger 3M 65'' touch-table at 4k resolution as our exhibit display to accommodate either 6 visitors interacting all around it or 3 visitors from one side.

To support an iterative development cycle, \textit{Sea of Genes} is web-based and written in ECMAScript 6, JavaScript 6.
JavaScript is lightweight and is suited for rapid prototyping. 
Each microbe had its own custom sprite and a set of animated behaviors derived from their transcripts.
The transcripts were functionally annotated, and patterns of sequential function were grouped into high level categories (e.g., genes involved in preparing a cell to divide, then genes involved in the actual division).
The time of expression for each transcript was determined by the time of day the sample was collected and the normalized amount of transcript in the sample.
Details on the visualization process are expanded upon in Section 5.
% Then each of these categories was given a sprite to represent that state and was either animated or had a particle effect.
% For example, the Sun harvester had four behaviors preparing, making sugar, releasing sugar, and dividing.
% To visually convey each behavior, they would have a unique image different from the default sprite image.
We provided a configuration file the museum staff can edit, allowing them to modify parameters such as the number of microbes and length of time for the 24hour period to cycle.
% Since the exhibit is written in a web-based language it can be deployed on any platform. 
To package for distribution we used the open source software Electron by github.io~\cite{Electron}.
For our own development we deployed the exhibit on OSX architecture.

\section{Design Considerations and Evaluation Methods}
Interpreting metagenomic data requires understanding microbes, their genes, and gene expression.
To create an experience around this complex dataset, we worked with C-MORE scientists to (1) synthesize their research into a narrative comprised of three related stories, and (2) apply techniques from established narrative frameworks to layer the following three stories into a cohesive narrative:
\begin{itemize}
    \item[\textbf{S1.}] \textbf{Microbial interactions occur in a predictable daily rhythm.} The first story conveyed to visitors that microbes form communities similar to larger organisms. The interactions and functions these microbes perform during the day and night differ.
    \item[\textbf{S2.}] \textbf{Genes turn on and off according to a daily rhythm.} The second story focused on how the microbial interactions and actions in the first story are a result of the gene expressions of specific genes, which control microbial function.
    \item[\textbf{S3.}] \textbf{Scientists collect data about the genes of microbes to make sense of the temporal patterns in microbial functions.} The final story was scientists collected data and identified the expressed genes responsible for a microbial function.
\end{itemize}
These stories were the synthesis of the C-MORE scientists' research and, taken together, could provide the public an understanding of how marine microbes are studied and what they do.
From discussion with the scientists we found S1 and S2 were closely related to one another, with S2 explaining the molecular underpinnings of, or genetic expression for, the behavior in S1.
S3 further elaborates that scientists study S2 to make sense of the temporal patterns in microbial interaction captured in S1.
In short, we needed to show the public (S1) microbes have interactions that occur in a predictable daily rhythm which (S2) are the result of gene expressions, and (S3) scientists analyze these gene expression data to identify temporal patterns.
How we present these stories is constrained by the considerations of an informal learning environment in an interactive science museum.
% We had to consider the museum exhibit experience itself in order to develop an exhibit that would effectively convey these three stories.

\subsection{Museum Considerations}
Museums are informal learning environments referred to as ``designed environments'' in which exhibits are developed to help structure visitor experiences, in line with institutional goals and values~\cite{national2009learning}.
In addition to facilitating visitor engagement and comprehension of complex datasets, the team needed to ensure the exhibit design considered the informal learning context.
% The museum context presents a set of considerations we must consider in our exhibit design in addition to facilitating visitor engagement and comprehension of complex datasets from unfamiliar domains.
The following considerations were identified and informed by our collaborators at the Exploratorium, which we used to constrain and guide our design process.
% These considerations include: (\textbf{C1}) the exhibit should exist in a free-choice learning environment, (\textbf{C2}) the design must accommodate learners with different interests and prior knowledge, and the exhibit should be (\textbf{C3}) readily decipherable and (\textbf{C4}) multi-user friendly.
% \begin{enumerate}   
%     \item free choice learning environment
%     \item diversity of learners
%     \item immediate apprehendability
% \end{enumerate}

\vspace{0.1in}

\noindent 
\textbf{C1. Free-choice learning environment.} 
As with other types of informal learning environments, the experiences in museums, as compared to the formal setting of the classroom, are motivated and guided by personal interests rather than compulsory requirements~\cite{FindingSig,Bedford}. 
This is often referred to as a ”free-choice” learning environment.
In such an environment visitors may not encounter or even choose to attend
to our exhibit.
The exhibit design must consider methods to attract and retain visitor attention.
As free-choice learning environments~\cite{national2009learning}, museums employ a variety of techniques to attract and sustain visitors interests and engagement at exhibits.
For example, DeepTree~\cite{Davis}, an interactive visualization of the tree of life had strategically placed features which invited attention and used the interactive table to encourage collaboration.
Likewise, the interactive plankton visualization in the Living Liquid project~\cite{JMa2012}, used an animated visualization paired with a tangible interface to captivate visitors’ interest and serve as gateway for exploration of plankton patterns.

\vspace{0.1in}

\noindent 
\textbf{C2. Public comprehension.}
The audience we design for is the visiting public, who are typically not domain experts.
% From \textbf{C1} although we are within a learning environment and seek to teach,
Since our museum attracts a diverse audience we do not know where, on the spectrum of novice to expert, our visitor's prior knowledge is.
Furthermore, the open space layout of the exhibits means there are no guarantees a visitor will come to the \textit{Sea of Genes} exhibit with the prerequisite knowledge learned from a prior exhibit~\cite{bell2009learning}.
Therefore, we cannot assume familiarity with the underlying dataset or domain itself. 
Similarly, we cannot assume representations which experts use for interpreting the data will translate to the public~\cite{Davis}.
However, we should be mindful to not trivialize the experience to exclude experts or people who want to explore the content matter deeply.

\vspace{0.1in}

\noindent 
\textbf{C3. Readily decipherable.}
When designing an exhibit in a science museum there is a need for fast decoding and ready interpretation of the visualization~\cite{ma2019decoding}.
In the Exploratorium’s \textit{Traits of Life} exhibit collection holding times at a signle exhibit ranged from 12 to 149 seconds~\cite{hein2003traits}.
In other words, visitors have a short dwell time at exhibits and within this time they need to decode what is visually presented.
Our design should thus accelerate this decode process.

\vspace{0.1in}

\noindent 
\textbf{C4. Support Multi-user Interaction.}
Exhibit design must allow for multiple visitors to view or interact. 
This comes in both the need to support social groups who frequent the museum and facilitate collaborative learning~\cite{falk2016museum}.
There are also logistical reasons for multi-user exhibits, such as preventing queues and facilitating visitor movement in the overall exhibit space, and providing more visitors access to an exhibit.
Designing for multiple users has several implications.
Because we cannot assume a visitor will come to the exhibit in its initial state, the design should ensure a visitor's can interpret and interact with the exhibit regardless of the state of the visualization.
Furthermore, a visitor's interaction with the exhibit must not adversely affect another visitor's experience.
Ideally, there are supports to encourage visitors to share their thoughts with each other and come to a common understanding of their shared exhibit experience.

\subsection{Evaluation Process}
Formative evaluation is an integral part of the iterative exhibit development process at the Exploratorium (Figure~\ref{fig:brainstorm} e).  
Depending on the complexity of the exhibit, development may entail several rounds of prototyping and evaluation, with each successive round testing prototypes with modifications informed by visitor feedback and behavior data collected through evaluation. 
For \textit{Sea of Genes} we conducted three successive rounds of prototyping and evaluation.
\section{Visualization Design}
% Sea of Genes was created using an iterative design process, shown in Figure~\ref{fig:stil}.  
% We sought to engage museum visitors with the stories found in a complex and unfamiliar dataset.  
Three iterations were designed and tested, each adding on one story from \textbf{S1--3}.  
During each iteration every design choice was guided by our considerations, \textbf{C1--4}. 
The following discussion is organized according to key design decisions made during our iterative development and evaluation process.

\begin{figure*}
	\includegraphics[width=\textwidth]{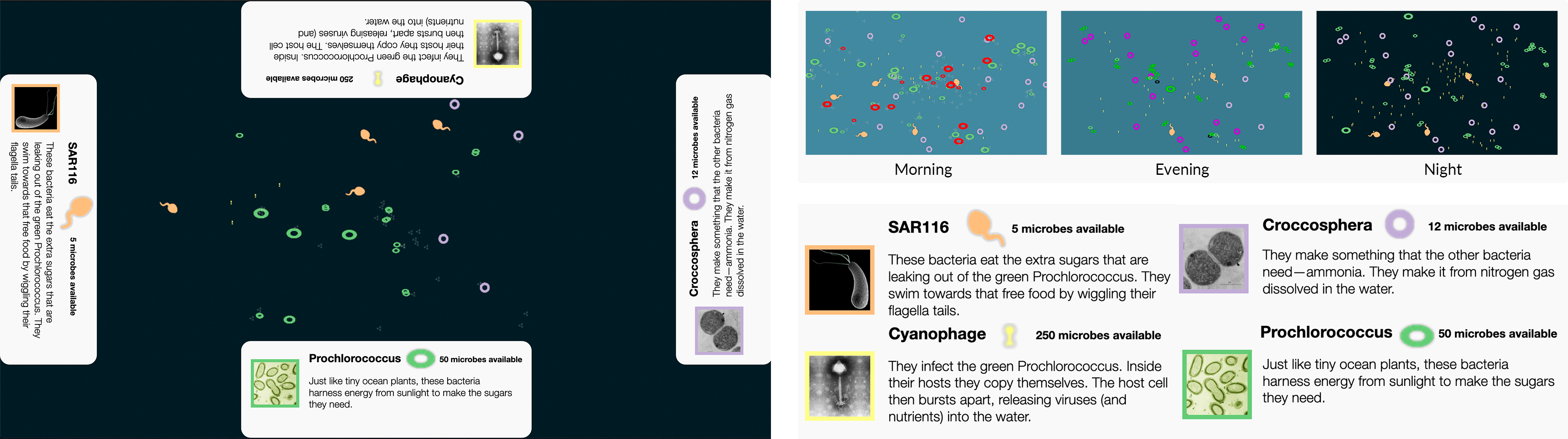}
  	\caption{(a) Sea of Genes Prototype 1 visualized four main microbial characters depicted as 4 unique icons.This visualization was tested on a large tabletop display. (b) Stills of the central animation at different times of day. (c) Cards providing information for each microbe. \vspace{-3mm}
%   	\vspace*{-0.12in}
  	}
	\label{fig:prototype-1}
% 	\vspace{-4mm}
\end{figure*}

\subsection{Constructing the Stories}
% As described earlier in our design criteria, we synthesized the data to form a layered story.
The first step we took, guided by Lee et al.'s~\cite{lee2015more} approach, was to spend time exploring the data and extracting data excerpts to use and support \textbf{S1 and S2}, as described in Section 4.
A study was conducted earlier at the Exploratorium to examine prior knowledge and interests in marine microbes and metagenomics. 
A large majority (96\%) of the 136 visitors interviewed described microbes by a role they believed microbes played, while few used scientific taxonomic classifications~\cite{ma_2011}.
Consequently, we decided to focus on functional roles.
% Prior studies informed us that visitors were unfamiliar with both how gene activity is analyzed and the microbes~\cite{ma_2011,lead2013next}.t
% We reduced the number of microbes we would focus on with the assumption it would improve visitor comprehension~\cite{hullman2011visualization}. 
To identify familiar functions from the dataset we referred to the Next Generation Science Standards (NGSS)~\cite{lead2013next} and consulted with our partners at University of Hawaii (UH).
The science standards specify science concepts taught at each grade level and are used to guide the design of educational experiences.
The Exploratorium often designs for middle school level; however, we found that most of the functions were not covered until high school.
With this in mind we consulted with our partners, who suggested selecting microbes from their dataset based on their roles in a microbial ecosystem and could demonstrate \textbf{S1--2}.
Selecting microbes based on roles rather than taxonomic classification would assist visitor familiarity~\textbf{C2}.
Four microbes were chosen for the first prototype (Figure~\ref{fig:brainstorm}). 
We selected phototrophs, \textit{Prochlorococcus}, which draw energy from the sun, heterotrophs, \textit{SAR116}, which consume other forms of energy like sugar, photo-heterotrophs, \textit{Crocosphaera}, which draw energy from the sun and eat other forms of energy, and viruses.
Viruses are not in NGSS; so, we relied on another prior study~\cite{spiegel2013engaging} that found \protect{71\%} of teens knew viruses caused infections, and \protect{79\%} recognized images of the type of virus used in the Hawaii dataset.
Next, we chose microbial functions to animate for each microbe. 
We selected functions from the data that had a strong daily pattern and were familiar to museum visitors~\cite{ma_2011}, selecting functions behind photosynthesis and cell division, which are concepts that are encountered in U.S. middle schools according to NGSS.

Prior studies~\cite{ma_2011,spiegel2013engaging} suggest visitors believed microbes had a larger role in our ecosystems.
However, we needed to determine which story to center the design around.
Our previous study~\cite{ma_2011} found 95\% of visitors knew microbes lived in the ocean, and although 28\% were initially surprised that microbes have genetic material, a majority (71\%) when told this fact found it reasonable and believable.
\textbf{S2 and S3} required explaining to the public the link between microbes and genetics.
Prior work~\cite{lanie2004exploring} found the general public had a limited understanding of basic genetic terms and concepts, suggesting that visitors would have difficulty with \textbf{S2} and hence \textbf{S3}.
For this reason the team decided that the main feature of the exhibit should be \textbf{S1}, an animation of microbial behavior with familiar descriptions.
% and as a way to show visitors microbes follow a daily pattern (\textbf{S1}).

\subsection{Prototype 1}
Our scientific partners at UH helped identify which stories could be told from their data.
One of the stories within the dataset was that microbes have function that are on a daily cycle.
Based on a previous study~\cite{ma_2011}, we focused on this daily cycle of microbial functions, which we predicted may give visitors a more familiar entry point in to the metagenomics data.

% After assembling our stories, the team focused on the form the visualization should take to address the design goals of the exhibit.  
% A previous study was conducted on Exploratorium visitors' understanding of microbes~\cite{ma_2011}.
% The study found that 95 percent of visitors knew microbes lived in the ocean, and that the majority (71 percent) were aware that microbes have genetic material.
% Furthermore, it found that museum visitors had an interest in the functions of microbes.
% From this study we could assume visitors knew microbes have genes, and focusing on functions would be attractive to visitors.

We collaborated with our scientific partners at the UH to create a visualization with somewhat familiar representations to visitors \textbf{(C2)} and to simplify the complexity of the data \textbf{(C3)}. 
From our discussions we created a model of microbial interactions that could best tell \textbf{S1} and offer some familiarity for visitors.
The model simulates a 24 hour period showing the functions each microbe performed during this time.
The objective of Prototype 1 was to see if visitors could follow \textbf{S1}. 
If they were able to do so then we would try to layer in \textbf{S2} and finally \textbf{S3}.

We chose to have a central animation based on prior work~\cite{chanlin1998animation} indicating that animations could be effective for conveying these concepts to novices,~\textbf{C2}.
However, animations, when designed for teaching those with varying domain knowledge, require varying the presentation forms to be effective in achieving a desired learning task~\cite{chanlin1998animation, kalyuga2008relative}.
We decided to layer the three related stories in an exhibit, we sought to present each story within a form, starting with~\textbf{S1} as an animation.
We chose \textbf{S1} to be the focus of the animation since of the three stories it could have the most familiarity with visitors~\cite{ma_2011} and, as the first story, it is the foundation upon which the other stories are built.
This animation would serve as our entry point~\cite{hornecker2007entry}, and be the centerpiece to attract visitors to the exhibit.
% We felt this was a valuable opportunity to evaluate and test the capabilities of animation when used for explaining metagenomics to the public.
This would also provide us with the opportunity to determine if a ``well-designed'' expository animation actually contains all the elements needed to draw the learners' attention to the right place at the right time~\cite{berney2016does} allowing visitors to quickly decode (\textbf{C3}) and understand \textbf{S1}.
%  as well as room for us to experiment with interaction \cite{kalyuga2008relative}
% We hypothesized that seeing the simulated behavior of the different organisms animated would help visitors make sense of the time-ordered interactions of a microbial ecosystem. 
% Explain each of these and why is it needed how does the elements play a role and why did
% we find them necessary to explain our stories.
% It depicted a twenty-four hour loop where visitors could observe behaviors for each microbe.
% For this first prototype we needed a device to engage visitors so we allowed them to add in additional microbes.
% Use of animation in p1
With Prototype 1 our intent was to have a minimalist animation.
Our animation (Figure~\ref{fig:prototype-1}a) was driven by our curated model of microbial interaction and portrayed microbes as icons interacting with one another. 
This animation had three main elements: (1) Four main microbial characters (Figure~\ref{fig:prototype-1}c), (2) A background which transitioned from light to dark blue (Figure~\ref{fig:prototype-1}b) and back over a set period of 45 seconds (\textbf{C3}) to illustrate a 24 hour period, and (3) Control panels that described each of the four microbes (Figure~\ref{fig:prototype-1}a). 
The control panels were static and only provided textual information about each microbe.
We included a simple visitor interaction of tapping on the control panel to inject viruses into the pool.
Our goal was to communicate \textbf{S1} using an animation that showed the functions of microbes during a 24 hour period informed by metagenomic data.
% We wanted the first story,~\textbf{S1}, to come across, i.e., for visitors to observe that microbes are interacting.

% \subsubsection{Takeaways and Next Steps}

The evaluation for Prototype 1 sought to determine if visitors could interpret this first story from the animation.
The exhibit was placed near other exhibits that focused on microorganisms.
One was an exhibit on the microbes that live in the termite gut, with a live microscope view of these microbes.
The other was a Winogradsky panel that shows microbial diversity and discusses energy production.
While there was not a designed exhibition, this context seemed to best support the content of \textit{Sea of Genes}.

To recruit participants for the evaluation, an evaluator stood near the exhibit and approached every third person as they walked passed a predetermined imaginary line near the exhibit.
% To recruit study participants, an evaluator stood near the exhibit and, of the visitors who appeared eight years or older, approached every third person as they walked passed a predetermined imaginary line near the exhibit.
If the systematically selected visitor was with a group, the whole group was invited to participate as well. 
Consenting visitors were asked to use the exhibit.
Because it was an early prototype and not all of the labels or the touch interactivity was implemented, the evaluator verbally described these aspects to the participants:
\begin{enumerate}[itemindent=2.90em]
    \item[Evaluator:] \textit{This exhibit shows how microscopic life behave in the ocean. The water changes color from dark, for night, to light blue, for day. You can release different organisms into the water to see what they do.  The touch is not working yet, so just let me know which organism you want, and I'll release them for you.}
\end{enumerate}

When participants indicated that they were done, the evaluator asked the visitor who interacted the most within their group a set of questions designed to gauge usability and comprehension. In this evaluation, we talked with a total of 38 visitors\footnote{The demographic breakdown of the evaluation participants was: 18 adults, 12 teenagers, and 8 children, with 20 females and 18 males.} over the course of four days. 
% For prototype 1, we used the evaluation process described in Section 4.2. 
% We were interested in points of confusion while interpreting the animation. 
% We also wanted to determine if the first story of our narrative could be conveyed via the animation.
% This evaluation was performed over four days. 
% We recruited and talked  
When we asked visitors what they found interesting, over 55\% of them talked about the interactions between microbes in general, with a minority (6 out of 38 visitors) mentioning a specific interaction shown in the animation, for example:
% We found that the majority of visitors noticed the microbial interactions and behaviors with a smaller minority noticing the day- night cycle.
% That is, 55\% of the 38 visitors reported interest in seeing interactions between microbes, while We x\% of visitors commented on the interactions noticed: 
% This led us to believe they understood some aspect of microbial behaviors. 
% We received comments of visitors interpreting or noticing interactions:
% \begin{tabular}{p{.45in}p{2.6in}}
% Visitor3: & \textit{[It was interesting] infection of Cyanophage to Prochlorococcus, and SAR116 eating. How it [infected Prochlorococcus] burst and a bunch came out.}\\
% Visitor37: & \textit{One thing was sugar eaters and sun harvesters working together [ was interesting]. [I] noticed the sugar eaters were going by sugar harvesters and right after the sugar harvesters were making sugar.}\\\\
% \end{tabular}
\begin{figure*}
  \includegraphics[width=\textwidth]{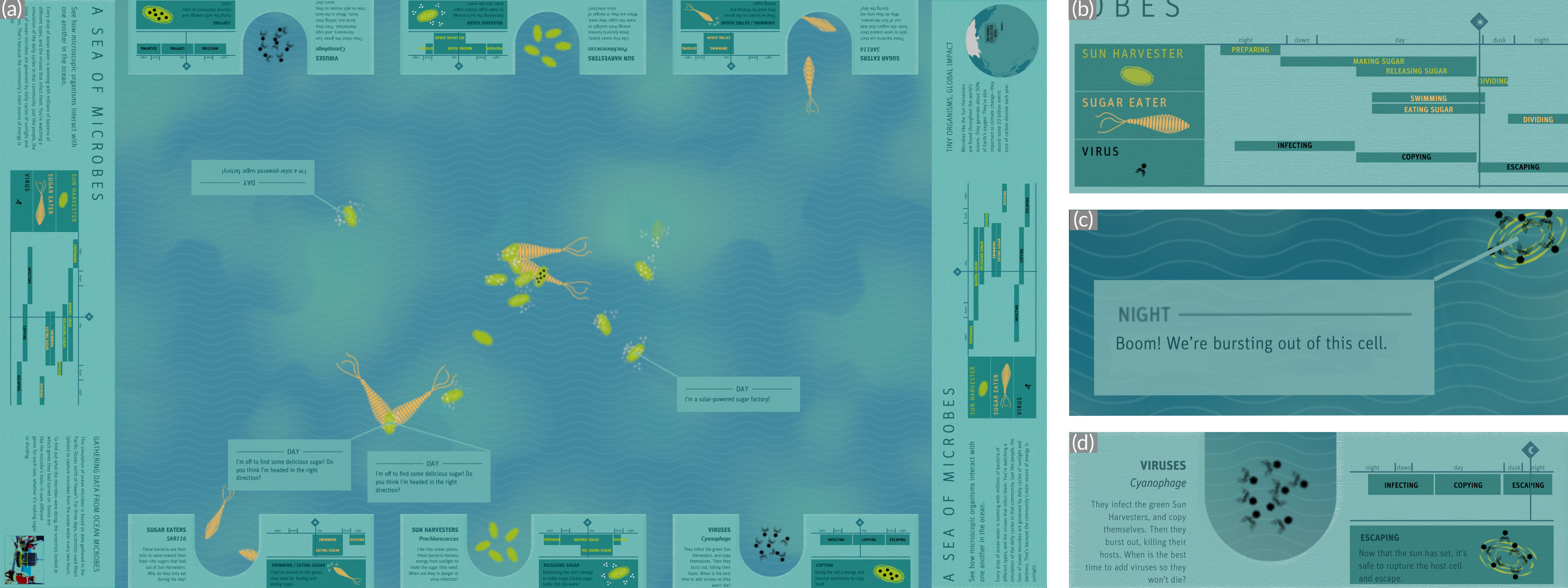}
  \caption{
  (a) Sea of genes Prototype 2's control panel and interpretive label. Panels are placed on two sides of the table to allow for more visitors to interact. Microbe annotations orient to the corresponding direction based on their position. (b) Timeline chart depicts an overview of what each microbe is currently doing and will do in the animation. (c) Annotation describes the microbe's current behavior. (d) Control panel describing the microbe and a timeline chart showing its current active behavior. \vspace*{-0.12in}}
  \label{fig:prototype-2}
%   \vspace{-5mm}
\end{figure*}

\begin{enumerate}[itemindent=2.55em]
\setlength{\topsep}{-0.03in}
\setlength{\itemsep}{-0.03in}
    \item[Visitor03:] \textit{[It was interesting] infection of Cyanophage to Prochlorococcus, and SAR116 eating. How it [infected Prochlorococcus] burst and a bunch came out.}
    \item[Visitor37:] \textit{One thing was SAR116 and Prochlorococcus working together [was interesting]. [I] noticed the SAR116 were going by Prochlorococcus and right after the Prochlorococcus were making sugar.}
\end{enumerate}

\noindent
% \begin{tabular} {p{.45in}p{2.6in}l}
% Visitor38: & \textit{The light and dark. Seeing
% difference between what's there and thrives in the light versus dark.}\\\\
% \end{tabular}
% However, 58\% of the visitors found parts of the animation confusing. 
% In particular, there were difficulties interpreting the animated behavior and making sense of that behavior; for example:
However, a majority (58\%) of the visitors found parts of the animation confusing. 
There were multiple reasons visitors had difficulty interpreting this animation.  
First, a few visitors complained that in the animation, the microbes' small size and thus poor resolution made it difficult to distinguish one type from another, e.g.: 
% \begin{tabular}{p{.45in}p{2.6in}l}
% Visitor1: & \textit{Some would disappear, especially this one [SAR116]. It was hard to see why they disappeared.}\\
% Visitor17: & \textit{These [sugar eaters] in particular appear to move randomly. A better graphic representation would be helpful.}\\ 
% Visitor24: & \textit{No idea what those guys [ammonia makers] do.} \\
% Visitor29: & \textit{Didn't understand what sugar eaters were doing.}\\\\
% \end{tabular}
\begin{enumerate}[itemindent=2.55em]
\setlength{\topsep}{-0.03in}
\setlength{\itemsep}{-0.03in}
    \item[Visitor01:]\textit{Some stuff was really small, so you couldn't see what was happening}
    \item[Visitor17:]\textit{These [Prochlorococcus] in particular appear to move randomly. A better graphic representation would be helpful.}
\end{enumerate}
\noindent Second, visitors could not decipher parts of the animation to make sense of microbial behavior, e.g.,:
\begin{enumerate}[itemindent=2.55em]
\setlength{\topsep}{-0.03in}
\setlength{\itemsep}{-0.03in}
    \item[Visitor24:]\textit{No idea what those guys [Crocosphaera] do.}
    \item[Visitor29:]\textit{Didn't understand what SAR116 were doing.}
\end{enumerate}
\noindent This was particularly the case when microbes appeared and disappeared as they were born and died, e.g.:
\begin{enumerate}[itemindent=2.55em]
\setlength{\topsep}{-0.03in}
\setlength{\itemsep}{-0.03in}
    \item[Visitor01:]\textit{Some would disappear, especially this one [SAR116]. It was hard to see why they disappeared.}
\end{enumerate}
\noindent A smaller number (29\%) of visitors noticed differences between day and night, even though the evaluator described the transition at the beginning of their exhibit use.
\begin{enumerate}[itemindent=2.55em]
    \item[Visitor38:]\textit{The light and dark. Seeing difference between what's there and thrives in the light versus dark.}
\end{enumerate}

\noindent
% The majority of visitors noticed the microbial interactions and behaviors with a smaller minority noticing the day-night cycle. 
Our evaluation of this prototype showed visitors were interested in the interactions between microbes.
A few noticed behaviors such as infection and eating. 
The majority, however, could only glean that there is a microbial community but may not have discerned the specifics of the behaviors or relationships.
% For some visitors, the noise in the form of random movement and poor assets hampered their ability to see the daily patterns encoded in microbial behavior.
% Our evaluation of this prototype showed visitors had interest and noticed interactions between microbes.
% A few were able to notice behaviors such as infection, eating, and working together.
% The majority, however could only glean that there is a microbial community here but failed to discern the specifics of the interaction.
This may have been due to the presence of too many unique animated elements; Pylyshyn and Storm~\cite{pylyshyn1988tracking} showed people can only track up to 5 independent moving targets accurately.
Rather than notice the individual unique animations between microbes, the excessive amount of moving elements may have led to it being processed as one entity.
Thus, we hypothesize this prototype fell to the Gestalt principle of Common Fate~\cite{palmer1999vision}, which states humans perceive visual stimuli that move in the same speed or direction as parts of a single stimulus.
Furthermore, processing both this visual information and decoding what it means may have distracted visitors from paying attention to the background color change.
We needed to improve how we portrayed our microbes to make clear the interactions of interest and focus visitor attention.
% At this stage we were unsure about the appropriate level of abstraction and realism.
Yet, this evaluation indicated that the animation was able to convey that microbes interact with one another, enough that 55\% of visitors talked about it, the first aspect of \textbf{S1}; however, it drew too much attention, resulting in few visitors seeing or discussing the daily aspect.
% Also not all microbial behaviors were equally accessible. 
Based on the evaluation data findings, it was clear that we had to improve our animation to better support visitors' interpretation of the microbial functions and their daily rhythm.
% We decided to try and guide visitors to stories with text in the form of annotation.
% We decided to simplify or eliminate some elements of the exhibit that detracted from the stories of interest. 
% To reduce noise we removed a microbe, after consulting with our scientific partners.
\begin{figure*}
  	\centering
	\includegraphics[keepaspectratio, width=\textwidth]{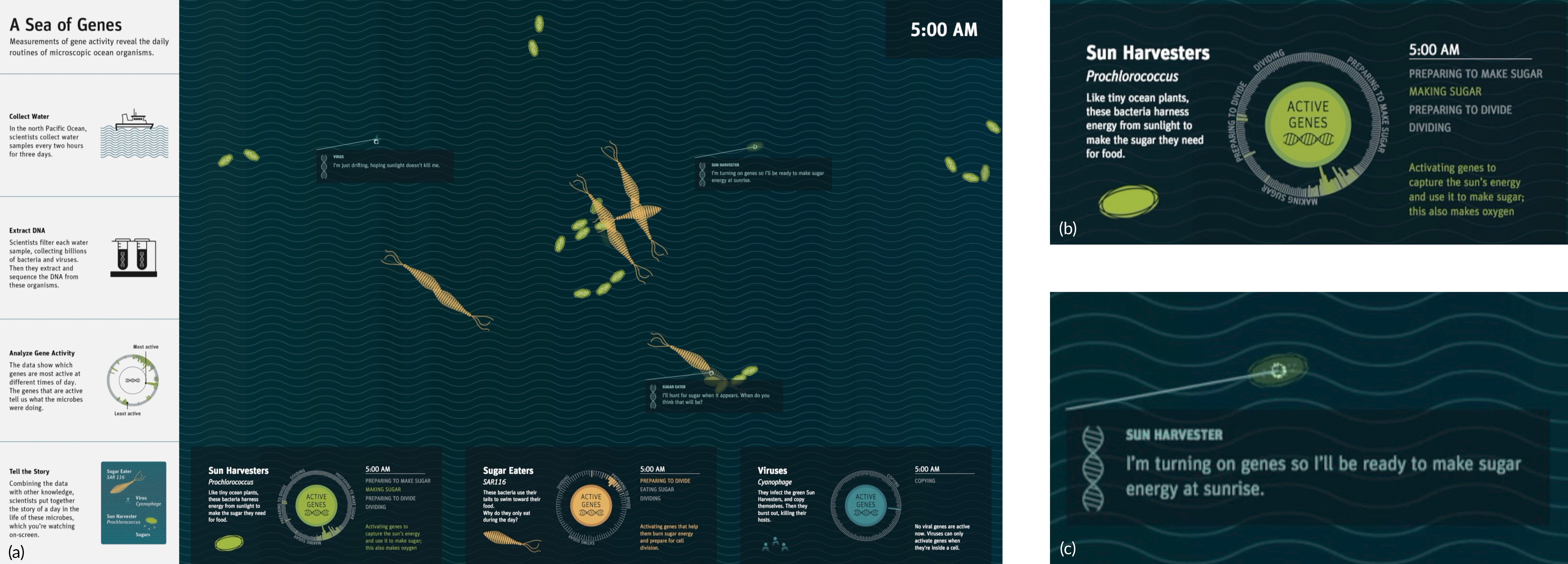}
  	\caption{(a) Final design of the Sea of Genes exhibit. (b) Legend containing information about Sun Harvesters and in the center is the activity gene showing the genes responsible for making sugar are being expressed.(c) Annotation showing what the Sun Harvester is doing.\vspace*{-0.12in}}
	\label{fig:p3}
% 	\vspace{-5mm}
\end{figure*}

\subsection{Prototype 2}
\begin{figure*}
  	\centering
	\includegraphics[keepaspectratio, width=\textwidth]{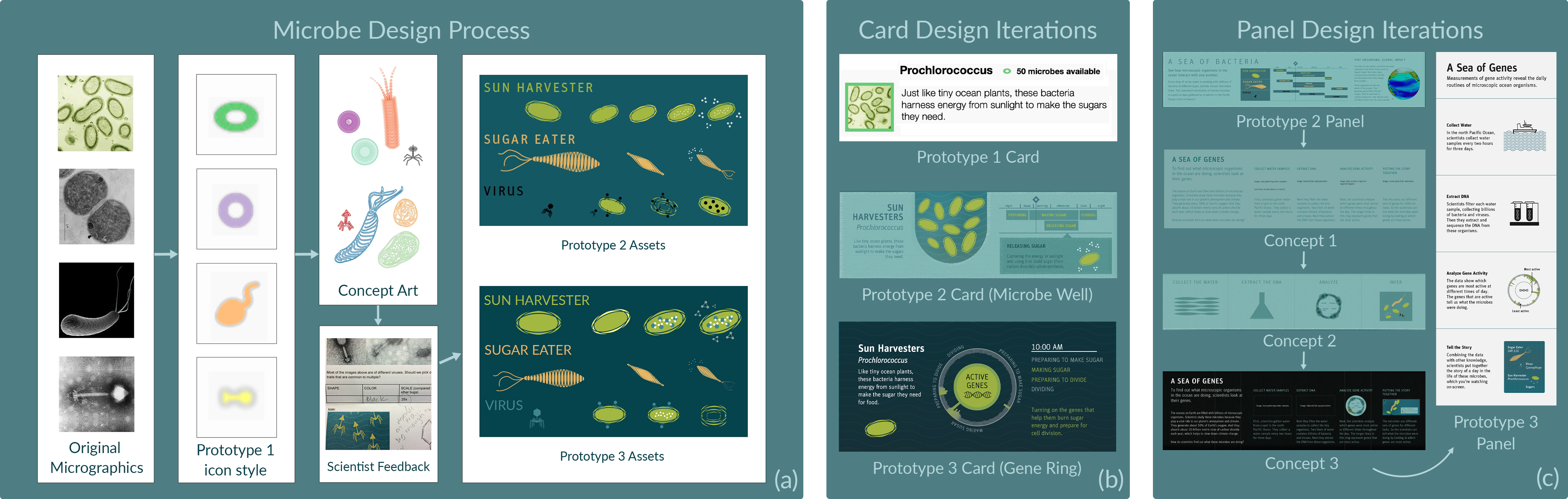}
  	\caption{(a) The design process of transforming the microbes into the characters of our stories over all prototypes. Initially we used icons to represent each, in the further iterations we used sprites. The final image depicts SAR116, Cyanophages, and Prochlorococcus in their final iterated state. (b) The card design changes between Prototype 1 to Prototype 3. (c) Various design iterations of the side panel from Prototype 2 to Prototype 3. \vspace*{-3mm}}
	\label{fig:microbes}
% 	\vspace{-5mm}
\end{figure*}

For the next version we wanted to (1) improve our animation presenting \textbf{S1} by making it easier for visitors to interpret and (2) layer on \textbf{S2} by visually communicating that gene expression going on and off is what drives the microbial functions seen in \textbf{S1}.
% direct visitors to the other stories, \textbf{S2}, of the narrative embedded in the visualization. 
% Also, we needed to better guide them such that they could focus on the details and discern the individual interactions between different microbial types as well as make connections to day and night cycles, \textbf{S1}.
% This would require modifying the animation.
Furthermore, for this version we elected a more designer-driven approach for presenting the narrative.
That is rather than let the visitors independently navigate the visualization and discover stories on their own, we sought to have more control on actively guiding visitors to the stories.
Because these changes would introduce more information to decode, we had to carefully revise the animation to convey the additional information without overwhelming the visitors~\cite{gershon2001storytelling,berney2016does}. 
To accomplish these tasks we focused on the following elements of the exhibit.
\begin{itemize}
\setlength{\topsep}{-0.03in}
\setlength{\itemsep}{-0.03in}
  \item Appearance of microbes and their behavior \textbf{(C1 and C3)}. Visual designers worked closely with the UH scientists and exhibit specialists to define how microbes and their behavior would appear in the exhibit (Figure~\ref{fig:microbes}). 
  \item Control panel for interaction and interpretation~\textbf{(C2 and C4)}. Central to interactivity was a control panel with a ``well'' of microbes that visitors could drag into the exhibit. The control panel also described the creatures and a timeline chart that tracked the timing of activities seen throughout the day (Figure~\ref{fig:prototype-2}a). 
  \item Annotations: New text was added to focus visitors to relevant information and reduce time finding what to observe~\textbf{(C3)}. These annotations also conveyed that microbial behavior and relationships follow a daily pattern (Figure \ref{fig:prototype-2}b). %~\ref{fig:microbes}c).
  
\end{itemize}
To improve interpretation of each story, we sought to reduce noise and confusion by both lowering the number of microbes and improving the quality of assets and animations~\cite{hullman2011visualization,tversky2002animation}.
Some visitors who saw Prototype 1 reported having trouble identifying the microbes.
Therefore, the microbe community was reduced from 4 to 3 and the assets of each microbe were changed to be more realistic compared to the previous version,~\textbf{(C3)}.
The simulated ocean animation now only contained three microbial types: SAR116, Cyanophage, and Prochlorococcus as shown in Figure \ref{fig:microbes}. 
We created non-scientific names for these microbial characters to reinforce their functional role in the ocean ecosystem and provide some familiarity to our visitors~\cite{mayer1998split,moreno2001case,ginns2005meta}.
``Sun Harvesters'' was the name given to Prochlorococcus, a microbe that makes energy from the sun.
``Sugar Eaters'' was the name given to SAR116, a microbe that lives on sugars produced by other microbes.
``The virus'' was the name given to cyanophage, a virus that infects Prochlorococcus.
% We wanted to visitors to build some connection or familiarity with these microbes that we felt was not possible in the previous iteration.
To reduce the amount of information visitors needed to process, we also showed a fewer number of microbes in the overall animation.

% \begin{figure}[ht]
%     \centering
% 	\includegraphics[width=\columnwidth]{Figures/microbe_dp.png}
%   	\caption{The design process of transforming the microbes into the characters of our stories over all prototypes. Initially we used icons to represent each, in the further iterations we used sprites. The final image depicts SAR116, Cyanophages, and Prochlorococcus in their final iterated state. \vspace*{-0.12in}}
% 	\label{fig:microbes}
% % 	\vspace{-6mm}
% \end{figure}

The previous prototype relied heavily on visitor participation and engagement; specifically, they had to invest time in interpreting and navigating the animation to discover \textbf{S1}.
This dependency we formed on visitor participation conflicts with \textbf{C1} so we chose to pivot in a different direction.
% We provided few devices to highlight or directly draw attention to \textbf{S1}.
In this prototype, we wanted to assert our narrative and lessen the time it takes to do so \textbf{(C3)}.
We added annotations to direct visitors to the stories \textbf{(C3)}.
Annotations have been used effectively in several studies of information visualizations~\cite{hullman2011visualization,heer2007voyagers, isenberg2007interactive,Groth2006} to add information, convey meaning, show data provenance, represent uncertainty, and highlight points of interest for users. 
Annotations also strengthen the narrative by drawing attention to aspects of the story we want to tell ~\textbf{(C3)}. 
Our annotations would pop-up and highlight a microbial action (Figure~\ref{fig:prototype-2}b) delivering a characterized message of what microbial function was occurring as well as reinforcing when it occurred.
This reinforcement aligns with the theories~\cite{chanlin1998animation,kalyuga2008relative} about the learning benefits of multiple forms of representation.
% Hart~/cite{hart2013human} The human lens: How anthropomorphic reasoning varies by product complexity and enhances personal value.
The characterized message was also designed to both anthropomorphize the microbe and highlight key interactions.
In marketing, anthropomorphism has been shown to have positive and significant influence on personal value~\cite{rauschnabel2014you}.
By providing human-like characteristics to the messages we theorized that visitors would engage more and process the narrative quicker \textbf{(C3)}.
These annotations would be triggered when an observable function occurred. 
Only one would be triggered at a time to not overwhelm the visitors~\textbf{(C3 and C4)} but to guide them through the story.
% The goal was for visitors to notice that over a 24 hour period different behaviors occur. 
% Fourth, we added a Timeline chart to each microbes card to reinforce the diel nature of their behaviors. 

Lastly, we updated the control panels (Figure~\ref{fig:prototype-2}a) by simplifying the text and providing a timeline chart to highlight the temporal aspect of microbial behavior.
This version included all microbes in the simulation and not just viruses.
The timeline was included to enable visitors to see the entire daily cycle for each type of microbe and provide context for what was occurring in the animation. 
An indicator, synced to the internal animation clock, would slide across the timeline to both reinforce the time and highlight what function each microbe was performing~\textbf{(C2 and C3)}.
% The introduction of an animated timeline chart synchronized to the period of the day was to reinforce that these behaviors happened at a specific time,
% It also showed the duration of an ongoing microbial behavior. 
% These devices were introduced to highlight the time dimension of microbial behavior.
Although animation implicitly illustrates time~\cite{ainsworth2008animations}, we needed to convey to the visitors the repetition of similar behaviors during the 24 hour period.
% To support \textbf{S3}, we added a side panel describing the data and why scientists are interested in it (Figure~\ref{fig:prototype-2}a). 
%Evaluation of prototype and Challenges with this process
%Stories we conveyed and stories we didn't. Why didn't they get across?
% \subsubsection{Takeaways and Next Steps}

The evaluation was conducted with 21 museum visitors recruited near the exhibit prototype\footnote{There were 15 adults, 5 teenagers, and 2 child participants.  
Nine were male, and 12 were female.}, following an evaluation protocol similar to that of Prototype 1. 
However, in this evaluation, the evaluator did not describe anything about the exhibit since all the exhibit labels and touch interactivity were implemented.
Instead, visitors were invited to use the prototype however they saw fit.

This evaluation found that visitors noticed the microbial interactions; when asked what they thought the exhibit was trying to show, 71\% mentioned microbial interactions, \textbf{S1}, with 86\% of visitors mentioning at least one microbial interaction when asked what they saw in the exhibit.
% Our evaluations showed that these improvements were noticed by our visitors.
% They continue to recognize community interaction and now could identify specific functions that occurred.
% We found that, similar to the first prototype, visitors understood that the exhibit was about marine microbes,~\textbf{S1}. 
% % The first story was conveyed.
% Majority of visitors (71\%) thought it was about the interaction of microbes, with 86\% of visitors mentioning at least one microbial interaction when asked what they saw in the exhibit. 
These findings suggest that a majority of the visitors understood aspects of the first layer of the story: Microbes interact with each other,~\textbf{S1}. 
But, they continued to struggle with noticing the daily cycle: Close to half of the visitor said that it was difficult to distinguish between day and night in the animation, and only one-third of the visitors mentioned a specific temporal pattern in microbial activities. 
% with only one-third of the visitors finding the day versus night differences interesting.
This was despite the addition of the timeline chart and emphasizing temporal patterns in both the control and the label.

The third layer of the story,~\textbf{S3}, was only partially conveyed. 
Although 62\% reported thinking that the animations was based on real data, a third of that group thought what they saw in the animated ocean was a representation of what researchers see. For example:
% A third of that group thought what they saw in the animated ocean was a representation of what researchers see.
% This is not the case.  
% For example:
% \begin{tabular} {p{.45in}p{2.6in}l}
% \\
%  Visitor8: & \textit{They went on a boat and collected
%  it in a bucket and put it in a petri dish and put it under a microscope and looked at it.}\\\\
% \end{tabular}
\begin{enumerate}[itemindent=2.55em]
    \item[Visitor08:] \textit{They went on a boat and collected it in a bucket and put it in a petri dish and put it under a microscope and looked at it.}
\end{enumerate}
% From this prototype, we learned that improvements we added improved the reception of the first story, presented via the central animation.
% We saw an improvement from 55\% to 71\% understanding the visualization depicts interactions of microbes.
% Furthermore, we saw 86\% visitors mention at least one microbial interaction that they saw.
These results suggest that the first story, depicted through the central animation, was communicated clearer due to the additions of anthropomorphized annotations and asset improvements.
We suspect that annotations helped visitors decode more readily, made the unfamiliar more familiar, and drew their attention to the salient parts of the visualization.
Visitors no longer needed to decipher what the role of ``procholoroccus'' was and instead could observe the ``Sun harvester'' explain simply what it was performing.

For our next version we sought to further improve how we convey \textbf{S2 and S3}.
Specifically, we sought to help visitors track temporal changes such as noticing microbes perform different abilities over a period, which many did not readily notice.
And, we needed to better highlight the underlying meta-genomics and meta-transcriptomics data,~\textbf{S2}.

\subsection{Prototype 3}
For the final iteration of \textit{Sea of Genes}, we focused efforts on sharpening the communication of \textbf{S2} and \textbf{S3} by emphasizing connections to the underlying metagenomic data.
% From the analysis in our first prototype we believe some of the same issues with visitors having difficulty and being distracted by all the various elements was still at play with our second prototype.
Our evaluation of Prototype 2 showed that although the prototype had fewer distractions relative to Prototype 1, \textbf{S2} and \textbf{S3} were largely unnoticed. 
The changes in Prototype 2 made the exhibit more effective in communicating \textbf{S1}.
% Based on the Apprehension and Congruence principles \cite{tversky2002animation} we hypothesized that the volume of microbes performing an action drew visitors attention much more than the animated background transition.  
% To make the stories stand out we reduced noise again.
% In short the interaction component itself did not help to convey any of the stories.
% The exhibit was now primarily an animation of microbes performing behaviors throughout a twenty four hour period.

We changed the orientation of the exhibit and reduced the number of panels around the table to focus attention on the animation. 
We hypothesized that having a single orientation for the exhibit would make it easier for visitors to decode the visualization \textbf{(C3)}.
Fixing the orientation appears to contradict our \textbf{C4}, supporting multi-user interaction. 
However, an evaluation of a similar tabletop visualization, Plankton Populations~\cite{JMa2012}, found visitors tended to use one side even though it supported multi-orientation use.
In line with \textbf{C2}, we moved the label to the left side of the exhibit and removed all complex graphic elements~\cite{mayer1998split,moreno2001case,ginns2005meta}.
The new label (Figure~\ref{fig:p3}a) told \textbf{S3}, expressing how the data was collected and how the representations in the visualization were linked in four steps: (1) Collect Water; (2) Extract DNA; (3) Analyze Gene Activity; (4) Tell the story.

% Prior prototype evaluations suggested visitors were learning new things about microbes in the ocean and how they interact.
% However, two of the three story layers were involved associating microbial interactions to gene expressions,~\textbf{S2,S3}. 
% These additional layers, that microbes behavior is driven by genes and that scientists collect data about their genes, needed to be emphasized further in the final version. 
% We revised and added a number of elements to reinforce these two stories.

% The majority of our efforts were focused on conveying that this exhibit was based on genetic data,.
To emphasize that the animation is based on genetic data, \textbf{S2}, we made the following additions: A large title that included the word genes (A Sea of Genes) (Figure~\ref{fig:p3}a), the legends and microbe annotations were adapted to refer to genes (Figure~\ref{fig:p3}b), and the iconic DNA helix was added to the middle of the gene activity wheel and the annotation boxes (Figure~\ref{fig:p3}c).
% We included the DNA helix icon on our annotations to emphasize the role of genetics .
We modified the design of the timeline chart to emphasize \textbf{S2} and passively support \textbf{S1}.
The new representation would display the underlying genomic data as a radial histogram (Figure~\ref{fig:p3}b).
We considered using a circos visualization, as they have been used by the New York Times to supplement stories on metagenomics and comparative genomics \cite{krzywinski2009circos}, however, it would only increase decode time (violating~\textbf{C3}) and we could not assume our visitors would be familiar with that representation~\textbf{(C2)}. 
Rather, we designed a visualization where each gene related to a particular behavior was grouped around the circumference of the circle (all of the photosynthesis-related genes under ``Preparing to Make Sugar''). 
Each gene had an activity range and was displayed as a dynamic histogram; length of the bar was determined by the normalized amount of transcript in the sample. 
So, through the course of 24 hours different areas of the circle would have waves of gene activity, similar to an equalizer.

Part of the success of conveying \textbf{S1} in the previous prototype may have been a result of the interactivity we provided visitors, i.e, the ability to add microbes to the animation.
% This is the same interaction we had in our first prototype, tapping on a card to add a microbe into the pool.
We chose to shift the story the interaction emphasized from \textbf{S1} to \textbf{S2}.
% which let visitors add microbes to the exhibit, which tied more to telling \textbf{S1}, to highlight genes.
% We suspected was that the interaction while it drew attention to behaviors, it prevented visitors from noticing the day to night transition and the other aspects of the exhibit.
% We suspected the interaction of adding a microbe narrowed visitors' attention 
% We theorized that visitors  were too focused on the specific microbe they introduced into the visualization.
Rather than adding a microbe, tapping on the revamped card (Figure~\ref{fig:p3}b) would highlight all microbes of that type and show their gene activity, Figure~\ref{fig:p3}c.

% \vspace*{-7mm}
% \subsubsection{Evaluation}

Prototype 3 was evaluated as part of a larger summative evaluation conducted by an external museum evaluation group, Inverness Research. 
This evaluation sought to find the key understandings that visitors came away with from their interaction with the exhibit.
While conducting the study, Inverness found that visitors were not spending sufficient time at the exhibit and therefore would not be able answer the questions in their exit interview.
So, they focused their efforts on mediated interviews, in which visiting groups were recruited to interact with the exhibit and answer questions immediately afterwards.
A total of 13 mediated interviews were conducted.
% Visitors time with the exhibit ranged from 2 to 89 seconds, with 53\% staying 10 seconds or less (30 out of the 56 observations).
The interview results reported  2 of the 13 groups understood that scientists collected data while on boats, and that something about what they collected is represented on the screen of the table.
However, there was no clear evidence that visitors understood that the data represented was gene expression.
Furthermore, the mediated interviews indicated that visitors could not figure out what to do or where to begin, and were often confused. 
For example, seven visitors said they were not sure where to start or what to do:
\vspace*{-3mm}

\begin{enumerate}[itemindent=2.55em]
    \item[Group A:]\textit{Honestly, its way over my head. I'm interested in what's happening, I just don't know what to do and I can't understand it. I guess there are Sun Harvesters and Sugar Eaters? There is a lot of empty space. I keep waiting for something to happen. My instinct is to ask how do I make it work so I can learn something? But I can't make it work. So there are three types of genes?}\vspace*{-3mm}
\end{enumerate}

Despite our efforts to design the exhibit to emphasize~\textbf{S2 and S3}, visitors were still extremely confused by what was presented.
The additions of annotations, improving the exhibit, changing to a single orientation, clearer animations, better assets, and fewer visual elements only seemed to help bring visitor attention to \textbf{S1}.
One reason, we believe, is usability problems made it difficult to convey the last two stories.
That is, visitors did not understand what their role was with the exhibit and did not feel they could participate with it.
The time spent figuring out their role \textbf{(C3)} resulted in visitors leaving the exhibit before learning anything deeper than aspects of \textbf{S1}.

\section{Discussion}
% This exhibit was developed and tested under a real scenario of museum exhibit design.
This project was a collaborative effort between several parties in an effort to develop a functional exhibit for deployment. 
With real-world collaborative projects, time and resources are often limited, which add significant constraints.
We worked with exhibit designers, who have a deep understanding of conveying scientific information to the public.
Our scientific partners at the University of Hawaii ensured the accuracy and fidelity of the data and their representation.
A design and data-visualization firm from the industry both expedited the process and offered their own insights into design and development.
Every prototype was tested with real targeted users.
The entirety of the project itself was built over the course of 2016--18.
From our significant evaluation effort over the iterative design process of developing \textit{Sea of Genes}, we identify relevant usability issues and areas of future work. 

With this paper, we introduced a set of considerations that should be addressed when designing narrative visualizations for an informal learning environment.
We believe applying these considerations, as we have shown in this paper, can support future designers attempting to visualize complex data for museum environments.
% dissecting an exhibit under our considerations, similar to how we have presented in this paper, is one that would help others examine the effectiveness of other structures and methods when communicating stories in such a setting.
At the time of development, we looked over literature~\cite{JMa2012,hsueh2016fostering,DeepTree,segel2010narrative,stolper2016emerging} for relevant techniques to apply to develop a successful exhibit under our considerations;~\textbf{(C1)} free-choice learning environment,~\textbf{(C2)} public comprehension,~\textbf{(C3)} readily decipherable, and~\textbf{(C4)} multi-user friendly.
% \begin{itemize}
% \setlength{\topsep}{-0.03in}
% \setlength{\itemsep}{-0.03in}
%     \item[\bf C1.]  Free-choice learning environment
%     \item[\bf C2.]  Public comprehension
%     \item[\bf C3.]  Readily decipherable
%     \item[\bf C4.]  Multi-user friendly
% \end{itemize}
We found limited research that met such an intersection and applied what we could, which gave some success.
With more time and resources we would have rigorously examined each technique in narrative visualization under these considerations.
However, in a real word setting of developing an exhibit where time and staffing are constrained, such analysis was not possible. 
We studied available narrative frameworks to identify concepts relevant for our work. We now present a discussion on the intersection of narrative visualization and museum design.

% What we find is this transitioning visitors from \textbf{S1} to \textbf{S2} under \textbf{C1-4} did not work.
% Given C1--4, we find that current narrative devices and storytelling frameworks---to first show microbial interactions and then explain their occurrence as a result of gene expressions---requires further exploration for the context of a museum.
% Our reflections on the limitations of this exhibit leads us to the following considerations;
% (A) Treat stories as graphs. (B) Use the context of stories as graphs to develop storytelling frameworks, and (C) Introduce a framework to support discussing stories in an informal learning environment.
% \vspace{0.1in}

\vspace*{0.05in}
\noindent
{\bf Storytelling Frameworks:} 
% As mentioned earlier in all three of our prototype evaluations, we see how changes to our visual encoding, noise reduction, asset improvements, and inclusion of annotations improved the public's ability to understand \textbf{S1}.
% These changes were guided by the literature~\cite{segel2010narrative,kosara2013storytelling,gershon2001storytelling,ma2011scientific,hullman2011visualization} and were successful in conveying that story but not the others.
% Limitation of Presentation Structure under Constraints
Segel and Heer~\cite{segel2010narrative} present a set of structures for balancing ``designer''-driven vs ``visitor''-driven narratives.
Stolper et al.~\cite{stolper2016emerging} provide an updated discussion of narrative visualization strategies with focus on systems with an ``author''-driven predefined narrative. 
These structures have been applied and shown to be effective in a variety of situations.
However, %we believe that 
these structures need to be viewed and evaluated under our considerations~\textbf{C1--4}.
Testing these structures and developing frameworks and methodologies that cater towards ``visitor-driven'' would aid the museum community.
Most analysis has been on settings (e.g. online journalism) where the user does not have distractions or free choice.
Visitors need to be engaged, and with other settings like the Exploratorium, exhibits need to support interaction and multiple users.
There is a clear need to aid exhibit designer's through further exploration and research into this space.

% Limitation of Martini Structure under Constraints
Here, we examine three structures: Martini Glass, Interactive Presentation, and Drill-down for our museum exhibit design considerations. 

The Martini Glass structure for narrative visualization allows for directing visitor attention explicitly to a set of points before releasing them with an understanding to make inferences for themselves.
However, with \textbf{C1} and evidenced by Boy et al.~\cite{boy2015storytelling}, it is difficult to assume they will get past the ``designer''-driven direct messaging and reach the exploration part.
A study~\cite{ma2013engaging} conducted at the Exploratorium examined if a narrative introduction could better contextualize the exhibit and found it had no real advantage over not-including it.
This introduction was a slideshow presenting where the dataset came from and its scientific significance, similar to \textbf{S3}.
Furthermore, under \textbf{C4} if the exhibit is at the exploration state, then new visitors are missing key information that allows them to truly participate, excluding them of this experience.

Next, the Interactive Presentation structure allows for an individual to progress through the story when they are ready to do so, and allows them to repeat steps.
% This structure has been shown to be successful in other contexts but in this context it can not work.
This structure, however, does not allow for multiple people to follow along~\textbf{(C4)}, in the sense of allowing any visitor to step forward or backwards, which could disrupt other's experience.
This constraint could be addressed by transforming the presentation into a looped animation.
The loop could also allow visitors to follow along back to points where they missed, ideally allowing for understanding at their pace.
% We also hypothesized that people would be intrigued \textbf{C1} by the animation and come to the exhibit.
However, even with using an animated loop, as the animation advances new information is continuously presented.
While visitors have started understanding a scene the animation has progressed, introducing new material to decode, leading to either confusion or frustration with the exhibit, as seen with Prototype 3.
%By the time visitors have started comprehending the scene it has changed, leading to frustration and confusion, seen with Prototype 3.

% Limitation of Drill down under Constraints
The Drill-Down approach appears to have the most promise but there is still a constraint on what you can train a visitor to do and expect~\textbf{C2}.
The structure allows for telling multiple related stories since it is built on drilling into sub-stories and adding new details.
However, there are several challenges with this structure when attempting to construct an exhibit experience with intuitive interactions that can be received by the majority of visitors.  There is the delicate balance of presenting either new visualizations or materials in the sub-views without overwhelming a visitor with content.  Then there is the additional challenge of ensuring that the exhibit is accessible to multiple visitors (i.e., if one visitor is drilling-down, it will not interfere with another visitor's experience). 
%The interactions, constructing an experience such that it is obvious to the majority of visitors (~\textbf{C4}) how to interact and decode these stories is a challenge.
Furthermore, depending on the relationships between stories this structure may struggle since it requires a central story to reach all other sub-stories.
Treating stories as graphs, documenting what kind of graphs each of these structures can present under \textbf{C1-4} is a direction that could merit great value for presenting in such conditions. 
How it handles such stories as \textit{Sea of Genes} with strong parallel themes between stories, \textbf{S1 and S2}, is an open question.

% How you contextualize the other stories under \textbf{C1-4} needs careful consideration and more work in this space could yield a new storytelling structure for informal learning environments.

\vspace*{0.05in}
\noindent
{\bf Stories as graphs:} 
Narratives are predominantly linear and are most effective in conveying a single perspective.
According to Spiro and Jehng~\cite{spiro1990cognitive},  \textit{``linearity of media is not a problem when the subject matter being taught is well structured and fairly simple.
However, as content increases in complexity and ill-structuredness, increasingly greater amounts of important information are lost with linear approaches.''}  
Can we create visual narratives that permit multiple perspectives and allow different narrative flows?
As Hullman and Diakopolus~\cite{hullman2011visualization} point out conveying a point of view requires careful over-emphasis.
It is well known that multiple perspectives are needed to learn complex topics~\cite{einsiedel2006challenges}.  
% This study also calls for research on narrative visualization. It highlights the importance of considering the motivation, target audience, and narrative structures when developing exhibits.  
We do not have good frameworks for classifying story complexity in a manner that can inform visualizations. 
A simple narrative has one causal pathway and is unidirectional. 
How do we characterize structures that are more complex? 
A taxonomy may allow us to develop visualization techniques. 
Effective storytelling is subject of interest to a diverse group of researchers in social sciences, computer science, and biological sciences.  
A taxonomy may allow us to map findings from these disparate domains and develop theories and guidelines. 
One approach could be to use graph theory. 
Here milestones, events, or information would be nodes and connections or flows represent edges.
A simple story is a unidirectional planar graph with no branches. 
Let us call this a basic graph.  
In our narrative, we had a more general network.   
The activities of an individual microbe (e.g., sun-harvester preparing sugar) is close to a basic graph. 
The activities of these microbes interacting (e.g., ``preparing sugar'' \& ``eating sugar'')  creates a more general graph.  
Since these connections occurred in parallel and all share the same time dynamics we have a directed graph that represents \textbf{S1}. 
The role of genes, \textbf{S2}, however, changes the structure of the graph. 
We could view the genomics as nested information. 
Embedded in each node corresponding to an event (e.g., ``preparing sugar''), there was genomic data (e.g., time of expression \&  amount of transcript). 
In our visualization, we present the embedded data in a narrative that was occurring in parallel in a separate space.
That is, we presented genomic data in a dynamic histogram on the bottom of the screen separate from the animation in the center while both update in parallel.
Our limited success in effectively connecting these two stories for the visitors highlights the need to consider other visualization techniques for these graphs. 
In short, we contend that there is a need for a richer taxonomy.

\section{Conclusion and Future Work}
% \noindent
Reflecting on this endeavor we find there is space for further research at the intersection of storytelling, data visualization, and informal learning.
% Sea of Genes presented us new challenges leading to new insights for future work.
The current storytelling structures are effective in many situations; however, delving deeper into the union of exhibit design and narrative visualization could extend the current structures, introduce "visitor"-driven methodologies, or offer adjustments and considerations to "author"-driven methodologies.

The process of communicating scientific findings as multiple stories visually in an informal learning environment brought many challenges.
We need better understanding of how to construct readily decipherable visual abstractions of a complex science, while maintaining scientific authenticity and accessibility to the public.
% Simple abstractions were engaging but it lacked scientific depth.
% Adding more scientific details diminished accessibility.
If the abstraction was too simple they didn't understand or notice the science, when the science was emphasized they were confused.
% In short, attempting to represent the data and inferences from the data with different representations did not work. 
% hinrichs2011, schmidt2007
% In future iterations of the exhibit, we would ensure the underlying connections between each of the stories is reflected visually by connecting representations.
Communicating multiple related stories is another challenge.
We need to ensure the underlying connections between each story is reflected visually. 
In our final iteration each story was embedded in a unique encoding; the community of microbes~\textbf{S1}, a radial histogram~\textbf{S2}, and a side panel~\textbf{S3}.
A visual link between these three was not explicit enough to be received by visitors.
We had some implicit clues, such as when a visitor touches a microbe a small radial histogram inside the microbe appears that correlates to the larger one.
At the time of development, designing a visual link was not considered. 
Our entry point into the exhibit was the community of microbes.
Yet, from the entry point to the two ancillary visualizations there isn't an explicit visual cue for a visitor to follow.
We lacked in our design a visual encoding that functions as a through-line for our stories.
In other words, there should be a visual encoding to reflect a common theme or consistent element within our stories.
Perhaps such an encoding would help visitors continue to the other stories beyond the entry point.
This requires the stories are not disjoint and have a minimum of one factor in common.
We reviewed our process for designing \textit{Sea of Genes}.
From our reflection we believe there is clear value in sharing experiences and lessons learned.
Overall this study supports retrospective analysis of design work in new cross-disciplinary domains even if the desired goals were not met.
Theories are typically shown to work in their documented
space; however, there is value in reporting how these theories behave
when tested outside of these documented spaces.
% Theories are typically shown to work in their documented space, however there is value in reporting how these theories behave when tested outside of these documented spaces.
% We believe there is great value in venturing into unknown spaces to identify open problems, provide lessons learned as well, as document case studies applying visualization techniques in these spaces. 
% Even if the desired goals are not reached documenting what was not effective or did not work is as valuable and useful as what is effective. for the many other designers creating visualizations of complex data for informal learning environments. 
We hope our extensive case study will stimulate additional research in approaches for visualizing complex data from unfamiliar domains for the public to explore in physical settings including museums and visitors centers.

% Designing an interactive visualization exhibit for the public is quite different from designing a visualization tool for domain scientists. 
% On one hand, scientists usually have informed routines and processes for navigating through their data. Visualization can leverage these processes and support them in their data analysis and discovery process. 
% On the other hand, visitors’ understanding of the domain and their approaches for deconstructing concepts are going to vary and more often than not will be unlike that of the scientists.  
% Another unique challenge in museum exhibit design is that there are several competing exhibits within each area of a given museum.
% Most visitors typically do not spend a great deal of time at one exhibit. 
% They usually approach an exhibit, look at the table or touch one button, then simply walk away after a glance. 
% Thus, design considerations for balancing between visual attractions, engagement for inquiry, amounts of information conveyed, and scientific authenticity are relevant but beyond the scope of this paper. 
% Furthermore, in our study, we did not evaluate how our design would serve multiple users, a topic pursued in our previous work.
% As big data plays an increasingly critical role in science, there is a corresponding need to provide learners with access to authentic, large-scale scientific data and to support their meaningful explorations. 

\section*{Acknowledgments}
The authors wish to thank Elisha Wood-Charlson and Ed DeLong for
sharing the SCOPE dataset and for their time and expertise. We would
also like to thank Eric Rodenbeck, Nicolette Hayes, and Andrew Wong
of Stamen Design for collaborating on the design and development of
the prototypes, and Meghan Kroning, Tamara Kubacki, Katherine Nam-
macher, and Janine Penticuff for evaluation assistance. This research
has been supported in part by the U.S. National Science Foundation
through grants DRL-1323214, DRL-1322828, and IIS-1528203 and by the Gordon and Betty Moore Foundation. Any opinions, findings, \
and conclusions or recommendations expressed in this material are 
those of the authors and do not necessarily reflect the view of 
the Foundations.

% The authors wish to thank Elisha Wood-Charlson and Ed DeLong for sharing the SCOPE dataset and for their time and expertise.
% % This exhibit would not have been possible without their assistance.
% %  and Leah Humphreys, Meghan Kroning, Tamara Kubacki, Katherine Nammacher, and Janine Penticuff for evaluation assistance.
% We would also like to thank Eric Rodenbeck, Nicolette Hayes, and Andrew Wong of Stamen Design for collaborating on the design and development of the prototypes, and Meghan Kroning, Tamara Kubacki, Katherine Nammacher, and Janine Penticuff for evaluation assistance.
% This research has been supported in part by the U.S. National Science Foundation through grants DRL-1323214 and IIS-1528203.
% This work was supported by the National Science Foundation under grants DRL-1322828, DRL-1323214 and IIS-1528203, and by the Gordon and Betty Moore Foundation.
% Any opinions, findings, and conclusions or recommendations expressed in this material are those of the authors and do not necessarily reflect the view of the Foundations.



\bibliographystyle{abbrv-doi-narrow}

% \nocite{*}
\bibliography{template}
\end{document}